\documentstyle[aps,epsfig]{revtex}
\newcommand{\be}{\begin{equation}}
\newcommand{\ee}{\end{equation}}
\newcommand{\bea}{\begin{eqnarray}}
\newcommand{\eea}{\end{eqnarray}}
\newcommand{\X}{{\vec X}}
\newcommand{\pro}{\partial}
\newcommand{\n}{\hat n}
\newcommand{\oneg}{\displaystyle\frac{1}{g}}

\newcommand{\D}{{\hat D}}

\newcommand{\vA}{{\vec A}}
\newcommand{\A}{{\vec A}}
\newcommand{\valpha}{{\vec \alpha}}

\newcommand{\hD}{{\hat D}}

\newcommand{\tD}{\tilde D}
\newcommand{\dfrac}{\displaystyle\frac}
\newcommand{\ba}{\begin{array}}
\newcommand{\ea}{\end{array}}

\newcommand{\nn}{\nonumber}

\begin{document}

\title{%
Dynamical Symmetry Breaking and Magnetic Confinement in QCD}

\author{%
    Y.M. Cho                  \\%% <== First author
{\it Department of Physics, College of Natural Sciences Seoul National University\\
Seoul 151-742, Korea\\
ymcho@yongmin.snu.ac.kr}\\
D.G. Pak  \\%% <== Second & Third author
{\it Asia Pacific Center for Theoretical Physics, Seoul 130-012, Korea\\
 dmipak@mail.apctp.org
}}

\maketitle

\section*{Abstract}

We present a gauge independent method to construct the effective action
of QCD, and calculate the one loop effective action of $SU(2)$ QCD
in an arbitrary constant background field. Our result establishes the
existence of a dynamical symmetry breaking by demonstrating
that the effective potential develops a unique and stable
vacuum made of the monopole condensation in one loop approximation.
This provides a strong evidence for the magnetic confinement of color
through the dual Meissner effect in the non-Abelian
gauge theory. The result is obtained by separating the topological degrees
which describe the non-Abelian monopoles from the dynamical degrees of
the gauge potential, and integrating out all the dynamical degrees of QCD.
We present three independent arguments to support our result.

\section{ Introduction}

One of the most outstanding problems
in theoretical physics is the confinement problem in QCD. It has
long been argued that the monopole condensation could explain the
confinement of color through the dual Meissner effect \cite{nambu,cho1}.
Indeed, if one assumes the monopole condensation, one could easily argue that
the ensuing dual Meissner effect guarantees the confinement \cite{cho2,ezawa}.
In this direction
there has been a remarkable progress in the lattice
simulation during the last decade.
In fact the recent numerical simulations have provided an unmistakable
evidence which supports the idea of the magnetic
confinement through the monopole
condensation \cite{kronfeld,stack}. Unfortunately so far there has been no
satisfactory field theoretic proof of the monopole condensation in QCD.
The purpose of this paper is to present a gauge independent method
to construct the effective action of QCD, and to establish
the magnetic confinement in the non-Abelian gauge theory
from the first principles of the quantum field theory \cite{cho3}.
{\it Utilizing a gauge independent parameterization of the gluon potential
which emphasizes its topological character
we establish the existence of the non-trivial vacuum
made of the monopole condensation in SU(2) QCD
in one loop approximation,
after integrating out all the dynamical degrees of the non-Abelian potential.
Our analysis shows that it is precisely the magnetic moment
interaction of the gluons which was responsible for the asymptotic
freedom that generates the monopole condensation in QCD}.
This strongly indicates that the magnetic confinement is indeed
the correct confinement mechanism of color in QCD.

To prove the magnetic confinement it is instructive for us to
remember how the magnetic flux is confined in the superconductor
through the Meissner effect.  In the macroscopic Ginzburg-Landau description
of superconductivity the Meissner effect is triggered by the
effective mass of the electromagnetic potential, which determines
the penetration (confinement) scale of the magnetic flux.
In the microscopic BCS description, this effective mass
is generated by the electron-pair (the Cooper pair) condensation.
This suggests that, for the confinement of the color electric flux,
one needs the condensation of the monopoles.
Equivalently, in the dual Ginzburg-Landau description, one needs the
dynamical generation of the effective mass for the
monopole potential. To demonstrate this one must first identify the
monopole potential, and separate it from the generic
QCD connection, in a gauge independent manner. This can be done
with an ``Abelian'' projection \cite{cho1,cho2},
which provides us a natural reparameterization
of the non-Abelian connection in terms of
the restricted connection (i.e., the dual potential) of the maximal Abelian
subgroup $H$ of the gauge group $G$ and the valence gluon (i.e.,
the gauge covariant vector field) of the remaining $G/H$ degrees.
With this separation one can show that
the monopole condensation takes place in one loop correction, after one
integrates out all the dynamical degrees of the non-Abelian gauge
potential.

The monopole condensation by itself does not
guarantee that it describes the true
vacuum of QCD. To prove that the monopole condensate
does indeed describe the true vacuum, one must calculate
the effective potential with an arbitrary background field configuration
and show that the monopole condensate becomes the absolute minimum
of the effective potential. In the following we prove that this
is indeed the case, at least in one loop approximation.
{\it We show that with an arbitrary background the color electric field creates an instability
to the effective action by generating an imaginary part.
This proves that the monopole condensation provides the only stable vacuum
of QCD which is unique. As importantly our analysis shows
that the gluon loop contributes
positively, but the quark loop contributes negatively,
to the imaginary part of the effective action.
This means that the gluons generate an anti-screening effect
by making pair annihilations, while the quarks generate a screening effect
by making pair creations}. This is a very important observation,
because this indicates that the gluons are not able to form
a hadronic bound state. A big mystery in hadron spectroscopy has been
the absence of the glueball states made of the valence gluons.
Our analysis provides a natural explanation why this is so.

\section{Abelian Projection and Extended QCD}

Consider $SU(2)$ QCD for simplicity.  A natural way to identify the
monopole potential is to introduce an isotriplet unit vector field
$\n$ which selects the ``Abelian'' direction (i.e., the color charge direction)
at each space-time point, and to
decompose the connection into the restricted
potential $\hat A_\mu$ which leaves $\n$
invariant and the valence gluon $\vec X_\mu$
which forms a covariant vector field \cite{cho1,cho2},
\bea
 & \vec{A}_\mu =A_\mu \n - \oneg \n\times\pro_\mu\n+\X_\mu\nonumber
         = \hat A_\mu + \X_\mu, \nn\\
 &  (\n^2 =1,~ \hat{n}\cdot\vec{X}_\mu=0),
\eea
where $
A_\mu = \n\cdot \vec A_\mu$
is the ``electric'' potential.
Notice that the restricted potential is precisely the connection which
leaves $\n$ invariant under the parallel transport,
\bea
\D_\mu \n = \pro_\mu \n + g {\hat A}_\mu \times \n = 0.
\eea
Under the infinitesimal gauge transformation
\bea
\delta \n = - \vec \alpha \times \n  \,,\,\,\,\,
\delta \A_\mu = \oneg  D_\mu \vec \alpha,
\eea
one has
\bea
&&\delta A_\mu = \oneg \n \cdot \pro_\mu \valpha,\,\,\,\
\delta \hat A_\mu = \oneg \D_\mu \valpha  ,  \nn \\
&&\hspace{1.2cm}\delta \X_\mu = - \valpha \times \X_\mu  .
\eea
This shows three things.  First the restricted potential
by itself forms an $SU(2)$ connection which
satisfies the full $SU(2)$ gauge degrees of freedom.
Moreover, $\hat{A}_\mu$ retains the full topological characteristics of the original non-Abelian potential.
Clearly the isolated singularities of $\hat{n}$ define $\pi_2(S^2)$
which describes the non-Abelian monopoles.  Indeed $\hat A_\mu$
with $A_\mu =0$ and $\hat n= \hat r$ describes precisely
the Wu-Yang monopole \cite{wu,cho4}.  Besides, with the $S^3$
compactification of $R^3$, $\hat{n}$ characterizes the
Hopf invariant $\pi_3(S^2)\simeq\pi_3(S^3)$ which describes the topologically distinct vacua
\cite{bpst,thooft}.  Secondly the valence gluon forms a
gauge covariant colored source of the restricted potential which does not
inherit any non-linear characters of the non-Abelian connection.
Finally this decomposition of the non-Abelian connection
is made without compromising the gauge invariance.
Obviously the decomposition holds in any gauge, and is gauge
independent.

The above discussion tells that $\hat{A}_\mu$ has a dual
structure.
Indeed the field strength made of the restricted potential is decomposed as
\begin{eqnarray}
&\hat{F}_{\mu\nu}=(F_{\mu\nu}+ H_{\mu\nu})\hat{n}\mbox{,}\nonumber\\
&F_{\mu\nu}=\partial_\mu A_{\nu}-\partial_{\nu}A_\mu,
~~~~H_{\mu\nu}=-\dfrac{1}{g} \hat{n}\cdot(\partial_\mu\hat{n}\times\partial_\nu\hat{n})
=\partial_\mu \tilde C_\nu-\partial_\nu \tilde C_\mu,
\end{eqnarray}
where $\tilde C_\mu$ is the ``magnetic'' potential
\cite{cho1,cho2}. This allows us to  identify the non-Abelian
monopole potential by
\bea
\vec C_\mu= -\frac{1}{g}\hat n \times \partial_\mu\hat n ,
\eea
in terms of which the magnetic field is expressed as
\bea
\vec H_{\mu\nu}=\partial_\mu \vec C_\nu-\partial_\nu \vec C_\mu+ g \vec
C_\mu \times \vec C_\nu
=-\frac{1}{g} \partial_\mu\hat{n}\times\partial_\nu\hat{n}=H_{\mu\nu}\hat n.
\eea
Notice that the magnetic field has a remarkable structure
\bea
H_{\mu\alpha}H_{\alpha\beta}H_{\beta\nu} =-\dfrac{1}{2}
H^2_{\alpha\beta}H_{\mu\nu},
\eea
which will be very useful for us in the following.

With (1) one has
\bea
\vec{F}_{\mu\nu}=\hat F_{\mu \nu} + \D _\mu \X_\nu -
\D_\nu \X_\mu + g\X_\mu \times \X_\nu,
\eea
so that the Yang-Mills Lagrangian is expressed as
\bea
{\cal L}=&-&\dfrac{1}{4} \vec F^2_{\mu \nu }=-\dfrac{1}{4}
{\hat F}_{\mu\nu}^2 -\dfrac{1}{4} ( \D_\mu \X_\nu -
\D_\nu \X_\mu)^2-\dfrac{g^2}{4} (\X_\mu \times \X_\nu)^2 \nn \\
&-&\dfrac{g}{2} {\hat F}_{\mu\nu} \cdot (\X_\mu \times \X_\nu).
\eea
This shows that the Yang-Mills theory can be viewed as
the restricted gauge theory made of the dual potential $\hat A_\mu$,
which has
the valence gluon $\vec X_\mu$ as its source \cite{cho1,cho2}.
But notice that here the valence gluon has the magnetic
moment interaction with the restricted potential.
This interaction plays the crucial role in
the monopole condensation as we will see in the following.

Obviously the theory
is invariant
under the gauge transformation (3) of the active type. But notice that
it is also invariant under the following gauge transformation
of the passive type,
\bea
\delta\hat{n}=0,~~~~\delta\vec{A}_{\mu}=\dfrac{1}{g}D_{\mu}\vec{\alpha},
\eea
under which one has
\bea
&\delta A_{\mu}=\dfrac{1}{g}\hat{n}\cdot D_{\mu}\vec{\alpha},~~~~~
\delta \vec{C}_\mu=0 , \nonumber\\
&\delta\vec{X}_{\mu}=\dfrac{1}{g}[D_{\mu}\vec{\alpha}
 -(\hat{n}\cdot D_{\mu}\vec{\alpha})\hat{n}]  .
\eea
This gauge invariance of the passive type plays an important role in the
background field method discussed  in the following.

\section{ Dynamical Symmetry Breaking and Monopole Condensation}

With this preparation
we will now show that the effective action of QCD, which
one obtains after integrating out all the dynamical degrees of the gluons from
the monopole background, can be written in one loop
approximation as
\bea
&~~{\cal L}_{g} = -\dfrac{Z}{4} \vec{H}^2_{\mu\nu}, \nn\\
&Z=1+\dfrac{22}{3}\dfrac{g^2}{(4\pi)^2}\Big(\ln\dfrac{gH}{\mu^2}-c\Big),
\eea
where $H=\sqrt{\vec{H}_{\mu\nu}^2}$,  $\mu$ is the modified minimal subtraction
parameter, and $c$ is a constant.
This generates the desired dynamical symmetry breaking and establishes the
magnetic condensation of the vacuum.

To derive the effective action consider the generating functional of (10)
\bea
W[J_\mu, {\vec J}_\mu]&=&
\int {\cal D}A_\mu {\cal D} \X_\mu  \exp \{ i \int
[-\dfrac{1}{4}{\hat F}_{\mu\nu}^2 -\dfrac{1}{4} ( \D_\mu \X_\nu -
\D_\nu \X_\mu)^2\nn \\
&-&\dfrac{g}{2} {\hat F}_{\mu\nu} \cdot (\X_\mu \times \X_\nu)
-\dfrac{g^2}{4}(\X_\mu \times \X_\nu)^2+A_\mu J_\mu
+ \X_\mu \cdot {\vec J}_\mu] d^4x\}.
\eea
We have to perform the functional integral
with a proper choice of a gauge, leaving $\vec C_\mu$ as a background.
To do this we first fix the gauge
with the condition
\bea
{\vec F}&=&\hat D_\mu (A_\mu \hat n + \vec X_\mu) =0 , \nn\\
{\cal L}_{gf}=&-& \dfrac{1}{2\xi}
\left[(\partial_\mu A_\mu)^2 + ({\hat D}_\mu \X_\mu)^2\right].
\eea
Notice that the gauge transformation of the passive type (11) plays the important role
in this background field method. With the above gauge fixing
the generating functional takes the following form,
\bea
W[J_\mu,{\vec J}_\mu]&=&\int {\cal D} A_\mu {\cal D}
\X_\mu {\cal D} \vec{c} {\cal D}\vec{c}^{~*}
\exp \{{~i \int[-\dfrac {1}{4}{\hat F}_{\mu \nu}^2} \nn \\
&-&\dfrac{1}{4} ( \D_\mu \X_\nu -\D_\nu \X_\mu)^2
-\dfrac{g}{2} {\hat F}_{\mu\nu} \cdot (\X_\mu \times \X_\nu)
-\dfrac{g^2}{4}(\X_\mu \times \X_\nu)^2\nn\\
&+&\vec{c}^{~*}\hat{D}_\mu D_\mu\vec{c}
-\frac{1}{2\xi}(\partial_\mu A_\mu)^2-\frac{1}{2\xi}
(\hat{D}_\mu\vec{X}_\mu)^2\nn\\
&+& A_\mu J_\mu+\X_\mu \cdot \vec J_\mu]d^4x\},
\eea
where $\vec c$ and ${\vec c}^{~*}$ are the ghost fields. In one loop
approximation the $A_\mu$ integration becomes trivial,
and the $\X_\mu$ and ghost integrations result in the
following functional determinants (with $\xi=1$),
\bea
&{\rm Det}^{-\frac{1}{2}} K_{\mu \nu}^{ab}\simeq
{\rm Det}^{-\frac{1}{2}}[-g_{\mu \nu}
 (\tilde D \tilde D)^{ab}
- 2gH_{\mu \nu}\epsilon^{abc} n^c], \nn\\
&{\rm Det} M^{ab}_{FP} \simeq {\rm Det}[- (\tilde{D}
\tilde{D})^{ab}],
\eea
where $\tilde D_\mu$ is defined with only the background $\vec{C}_\mu$.
One can simplify the determinant $K$
using the relation (8),
\bea
\ln {\rm Det}^{-\frac{1}{2}} K&=&\ln {\rm Det}[-(\tilde{D}\tilde{D})^{ab}]\nn\\
&-&\frac12\ln {\rm Det}[-(\tilde{D}\tilde{D})^{ab}
+i \sqrt{2}gH\epsilon^{abc}n^c]\nn\\
&-&\frac12\ln {\rm Det}[-(\tilde{D}\tilde{D})^{ab}-i \sqrt{2}gH\epsilon^{abc}n^c].
\eea
With this the one loop contribution of the functional
determinants to the effective action can be written as
\bea
&\Delta {S}=i \ln {\rm Det}[(-\tilde{D}^2 + \sqrt{2}gH)
               (-\tilde{D}^2 - \sqrt{2}gH)] ,
%&+\dfrac{i}{2} \ln {\rm Det}[(\tilde{D}_-\tilde{D}_- +\sqrt{2}gH)
%               (\tilde{D}_+\tilde{D}_+ -\sqrt{2}gH)],
\eea
where now $\tilde{D}_\mu$ acquires the following Abelian form,
\bea
\tilde{D}_{\mu} =\partial_\mu + ig\tilde{C}_\mu .\nn
\eea
Notice that the reason for this simplification is precisely
because our restricted potential ${\hat A}_\mu$ originates from the Abelian projection.

With this one can use the heat kernel method and
calculate the functional determinant.
For a covariantly constant $\vec H_{\mu\nu}$ one finds \cite{savv,ditt}
\bea
&\Delta{{\cal L}}=\dfrac{1}{16 \pi^2}\int_{0}^{\infty} \dfrac{ d t}{t^{2-\epsilon}}
 \dfrac{g  H/\sqrt{2} \mu^2  }{\sinh  (g H t/\sqrt{2}\mu^2 ) }\nn \\
&\times[ \exp (-\sqrt{2}g H t/\mu^2 )+  \exp (\sqrt{2}g H t/\mu^2 )],
\eea
where $\epsilon$ is the ultra-violet cut-off parameter.
The integral contains the (usual) ultra-violet divergence,
but notice that here it is also plagued by a severe
infra-red divergence. This, of course, is precisely
what one should have expected, because such an infra-red divergence
is an unavoidable characteristics of QCD. So the important issue now
becomes how to regularize the infra-red divergence.

To find the correct infra-red regularization, one must understand
the origin of the divergence. The infra-red divergence
can be traced back to the
magnetic moment interaction of the gluons that we have in (10), which
is also well-known to be responsible for the asymptotic
freedom \cite{gross}. This magnetic interaction
generates negative eigenvalues in Det K in the long distance
region, which cause the infra-red divergence.
More precisely when the momentum $k$ of the gluon
parallel to the background magnetic field becomes smaller than
the background field strength (i.e., when $k^2 < gH/\sqrt{2}$), the lowest
Landau level gluon eigenfunction whose spin is parallel to
the magnetic field acquires an imaginary energy and thus becomes tachyonic.
It is these unphysical tachyonic states which cause the infra-red
divergence. So one must exclude these tachyonic modes in the calculation
of the effective action, when one makes a proper infra-red regularization.
Including the tachyons in the physical spectrum will surely destablize
QCD and make it ill-defined.

With this understanding we can do both the
ultra-violet and infra-red regularizations simultaneously with due care.
Excluding the contribution of the unphysical modes
we have  \cite{cho3}
\bea
&\Delta{{\cal L}}=\dfrac{11g^2}{96\pi^2} H^2
(\dfrac{1}{\epsilon}-\gamma)-\dfrac{11g^2}{96\pi^2}H^2(\ln
\dfrac{gH}{\mu^2}-c_1)\nn\\
&c_1=1-\dfrac{1}{2}\ln2-\dfrac{24}{11}\zeta'(-1,\dfrac{3}{2})=1.29214... ,
\eea
where $\gamma$ is the Euler's constant and $\zeta'(x,y)$ is
the generalized Hurwitz zeta-function.
From this we finally obtain (with the modified
minimal subtraction)
\bea
{\cal L}_{g}=-\dfrac{1}{4}H^2 -\dfrac{11g^2}{96\pi^2}H^2(\ln
\dfrac{gH}{\mu^2}-c_1).
\eea
This completes the derivation of the effective action (13).

Clearly the effective action provides the following non-trivial
effective potential
\bea
V=\frac{g^2}{4}(\vec{C}_\mu\times\vec{C}_\nu)^2\Big \{ 1
+\frac{22}{3}\frac{g^2}{(4\pi)^2}
\Big[ \ln \frac{g^2[(\vec{C}_\mu\times\vec{C}_\nu)^2]^{1/2}}
{\mu^2}-c_1 \Big]\Big\},
\eea
which generates the desired monopole condensation of the vacuum,
\bea
<H>=\frac{\mu^2}{g} \exp\Big(-\frac{24\pi^2}{11g^2}-\frac12+c_1 \Big).
\eea
Notice that with $\alpha_s = 1$ we have
\bea
    \dfrac{<H>}{\mu^2} = 0.11225... .
\eea
The vacuum generates an ``effective mass'' for $\vec C_\mu$,
\bea
m^2=\frac{11g^4}{96\pi^2}\Big<\frac{(\vec{C}_\mu\times
\vec{H}_{\mu\nu})^2}{H^2}\Big>,
\eea
which demonstrates that the monopole condensation indeed generates the
mass gap necessary for the dual Meissner effect. Obviously the mass scale
sets the confinement scale.

To check the consistency of our result with the perturbative QCD
we now discuss the running coupling and the renormalization.
For this we define the running coupling $\bar g$ by
\bea
\frac{\partial^2V}{\partial H^2}\Big|_{H=\bar H} =\frac{1}{2}\frac{g^2}{ \bar g^2}.
\eea
So with $g \bar H = \bar\mu^2 \exp(c_1 - 3/2)$
we obtain the following $\beta$-function,
\bea
\frac{1}{\bar g^2} =
\frac{1}{g^2}+\frac{11}{12\pi^2}\ln\frac{\bar\mu}{\mu},~~~
\beta(\bar\mu)=-\frac{11}{24\pi^2} \bar g^3~,
\eea
which exactly coincides with the well-known asymptotic
freedom result \cite {gross}.
This confirms that the
asymptotic freedom and the monopole condensation have exactly the same
underlying dynamics.

In terms of the running coupling the renormalized potential is given by
\bea
V_{\rm ren}=\frac14 H^2\Big[1+\frac{22}{3}\frac{\bar g^2}{(4\pi)^2}
(\ln\frac{\bar g H}{\bar\mu^2}-c_1)\Big],
\eea
and the Callan-Symanzik equation
\bea
\Big(\bar\mu\frac{\partial}{\partial \bar\mu}+\beta\frac{\partial}{\partial \bar g}
-\gamma( \vec C_\mu)\vec{C}_\mu\frac{\partial}{\partial\vec{C}_\mu} \Big)V_{\rm ren}=0
\eea
gives the following anomalous dimension for $\vec C_\mu$,
\bea
\gamma( \vec C_\mu)=-\frac{11}{24\pi^2}\bar g^2.
\eea
This should be compared with that of the gluon field in perturbative QCD,
$\gamma(\vec{A}_\mu)=5\bar g^2/24\pi^2$ for $SU(2)$.

There have been many attempts to construct the effective action of QCD
in the literature, and in the appearance our vacuum (24)
looks very much like the old Savvidy-Nielsen-Olesen vacuum \cite{savv,ditt}.
But it must be emphasized that there are fundamental differences between the
earlier attempts and the present approach.
The earlier attempts had two problems.
First the separation between
the classical background and the quantum field was not gauge independent,
which made the one loop vacuum neither gauge invariant nor Lorentz
invariant. This violation of the gauge
invariance was of course a serious defect, but perhaps the more serious
problem was that the infra-red divergence was not properly regularized
in many of the earlier attempts. Indeed
it has been asserted that the Savvidy-Nielsen-Olesen
vacuum should be unstable, because the effective action
which defines the vacuum develops an imaginary part
\cite{savv,ditt},
\bea
Im \thinspace {\cal L}_{g} \Big |_{SNO} = ~\dfrac {g^2} {16\pi} H^2,
\eea
which destabilizes the vacuum through the pair creation of gluons.
This assertion of the instability of the
Savvidy-Nielsen-Olesen vacuum, which comes from
improper infra-red regularizations, has been widely accepted and
never been convincingly revoked.
Obviously, without a proper infra-red
regularization one can not expect to obtain the correct effective action
of QCD. Because of these defects the earlier attempts have not been
so successful.

{\it In contrast in our approach the separation of the monopole background
from the quantum fluctuation is clearly gauge independent.
Moreover our infra-red regularization
generates no imaginary part in the effective action. Because of these
we obtain a stable vacuum made of monopole condensation which is
both gauge and Lorentz invariant}.
Notice that the infra-red regularization in (20) is not just to remove the
infra-red divergence (there are infinitely many ways to do this). The infra-red
divergence that we face here in QCD is also different from those one encounters
in the massless QED. The infra-red divergence in the massless QED
comes from the zero modes. But these zero modes are physical modes,
which should not be excluded in the calculation of the effective
action. On the other hand the infra-red divergence
that we have here  comes from the unphysical modes,
and one must exclude these unphysical modes from the physical spectrum with
a proper infra-red regularization (Notice that in the earlier attempts
these tachyonic modes are incorrectly identified as
the ``unstable'' modes, but we emphasize
that they are not just unstable but unphysical).
And it is precisely these unphysical modes
that generate the controversial imaginary part in
the Savvidy-Nielsen-Olesen action.
So with the exclusion of the unphysical modes the instability
of the vacuum disappears completely.

As importantly in our approach we can really
claim that the magnetic condensation is a gauge independent phenomenon.
Furthermore here we have demonstrated that it is precisely
the Wu-Yang monopole that is responsible for the
condensation. Notice that in the earlier attempts it has never been clear
what was the source of the condensation, nor has it been simple
to show that the condensation
is indeed a gauge independent phenomenon.

Clearly  the quark loop makes an additional
contribution to the effective action. As we will see in the
following we have for the massless quarks,
\bea
&\Delta {\cal L}_{q}= \dfrac{g^2 H^2}{96 \pi^2}N_f(\ln \dfrac{gH}{\mu^2}-c_2),\nn \\
&c_2 =\gamma +\dfrac{1}{2}\ln 2 +\ln (2\pi) +\dfrac{6}{\pi^2}\sum_{n=1}^{\infty}
\dfrac{1}{n^2}\ln n  = 3.33163... ,
\eea
where $N_f$ is the number of the flavors of the quarks. With this
we obtain the following total effective action of $SU(2)$ QCD for the pure
magnetic background,
\bea
{{\cal L}}_{eff}&=&-\dfrac{1}{4} H^2-\dfrac{11-N_f}{96\pi^2}g^2 H^2
(\ln\dfrac{gH}{\mu^2}-c_3),\nn \\
&c_3& = \dfrac{11}{11-N_f} c_1 -\dfrac{N_f}{11-N_f} c_2.
\eea
The corresponding effective potential is plotted in Fig.1,
where we have assumed $\alpha_s = 1, ~\mu =1$, and $N_f = 2$. The effective potential clearly shows that there is indeed a dynamical
symmetry breaking in QCD.
\begin{figure}[tbp]
\begin{center}
\epsfig{file= 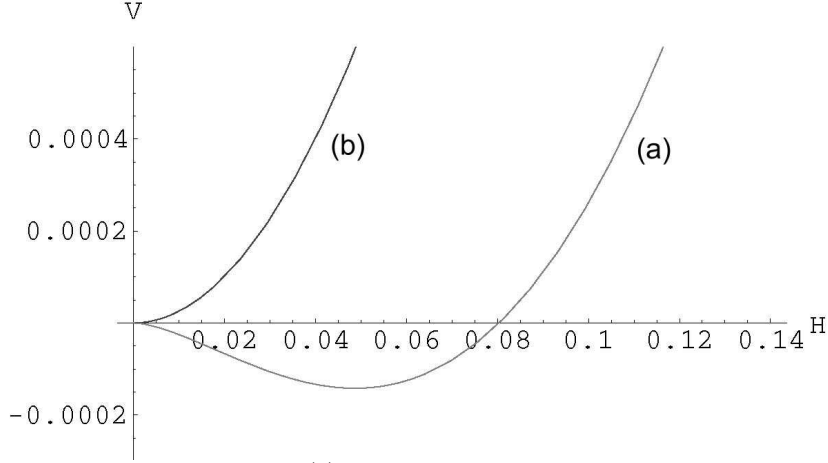, width = 11cm}
\caption[]{The effective potential of SU(2) QCD in the pure magnetic background.
Here (a) is the effective potential and (b) is the classical potential.}
\end{center}
\end{figure}

\section {Electric Background}

To make sure that our infra-red regularization is indeed the correct one
it is necessary to have an independent confirmation of the above result.
To do this it is instructive to calculate the effective action
with a pure electric background first.

So let us consider the general case where
the background field strengh $\hat F_{\mu\nu}$
contains both electric and magnetic components.
In this case we have
\bea
\hat F_{\mu\nu}= G_{\mu\nu}\hat n ,~~G_{\mu\nu} = F_{\mu\nu} +H_{\mu\nu},\nn
\eea
and the functional determinants of the gluon and the ghost loops are generalized to
\bea
&{\rm Det}^{-\frac{1}{2}} K_{\mu \nu}^{ab}\simeq {\rm Det}^{-\frac{1}{2}}[-g_{\mu \nu}
 (\hD \hD)^{ab}- 2g G_{\mu \nu}\epsilon^{abc} n^c],\nn \\
& {\rm Det} M_{FP} = {\rm Det} [-(\hat D \hat D)^{ab}] ,
\eea
where now ${\hat D}_\mu$ is defined with an arbitrary
background field ${\hat A}_\mu$.
Using the relation
\bea
G_{\mu \alpha} G_{\alpha \beta} G_{\beta \nu} = -\dfrac{1}{2} G^2 G_{\mu \nu}
-\dfrac{1}{2}(G \tilde G) {\tilde G}_{\mu \nu}
~~~~({\tilde G}_{\mu \nu}=\dfrac{1}{2}{\epsilon}_{\mu\nu\rho\sigma}G_{\rho\sigma}),
\eea
one can simplify the functional determinants of the gluon and the quark loops
as follows,
\bea
& \ln {\rm Det}^{-\frac 1 2}[(- g_{\mu \nu} (\hat D \hat D)^{ab}-2g G_{\mu\nu} \epsilon ^{abc} n^c] = \nn\\
& \ln {\rm Det} [(-\tD^2+2a)(-\tD^2-2a)(-\tD^2-2ib)(-\tD^2+2ib)],\nn\\
& \ln {\rm Det}M_{FP} = 2\ln {\rm Det}(-\tD^2),
\eea
where
\bea
a = \dfrac{g}{2} \sqrt {\sqrt {G^4 + (G \tilde G)^2} + G^2}, \,\,\,\,
b = \dfrac{g}{2} \sqrt {\sqrt {G^4 + (G \tilde G)^2} - G^2},
\eea
and now  $\tilde D_\mu$ is defined with an arbitrary background $A_\mu + {\tilde C}_\mu$,
\bea
\tD_\mu = \pro +ig(A_\mu + {\tilde C}_\mu).\nn
\eea
So for a pure electric background (i.e., for $a=0$) we have
\bea
\Delta {\cal L}  =  \dfrac{1}{16 \pi^2}  \int_{0}^{\infty} \dfrac{ d t}{t^{2-\epsilon}}
\dfrac{b}{\sin (bt) }[\exp(2ibt)+\exp(-2ibt)].
\eea
Notice that (unlike the pure magnetic background) the integrand
of the above integral has poles on the real axis, so that we must
specify the contour of the integral. Here the causality requires
the contour to pass above the real axis.

There are different ways to evaluate the integral, but a simple
and nice way of doing this follows from the observation that in
the imaginary time (i.e., in the Minkowski time) the role of
the electric and magnetic fields are reversed. So with the Wick rotation
of the proper time $t$ to the imaginary time $it$, the above integral acquires
the same form as (20). Indeed with the Wick rotation (39) becomes
\bea
\Delta {\cal L}  \longrightarrow  -\dfrac{1}{16 \pi^2} i^{\epsilon}  \int_{0}^{\infty} \dfrac{ d t}{t^{2-\epsilon}}
\dfrac{b}{\sinh(bt) }[\exp(-2bt)+\exp(2bt)].
\eea
From this we obtain
\bea
&\Delta{{\cal L}}=
-\dfrac{11b^2}{48\pi^2}(\dfrac{1}{\epsilon}-\gamma)
+\dfrac{11b^2}{48\pi^2}(\ln\dfrac{b}{\mu^2}-c_g)
-i\dfrac{11b^2}{96\pi},\nn\\
& c_g=1-\ln 2 -\dfrac {24}{11} \zeta'(-1, \frac{3}{2})=0.94556... .
\eea
So with the modified minimal subtraction we have (with the pure
electric background)
\bea
{\cal L}_{g} = \dfrac{b^2}{2g^2} +\dfrac{11b^2}{48\pi^2}
(\ln \dfrac{b}{\mu^2}-c_g)-i\dfrac{11b^2}{96\pi}.
\eea
It must be emphasized that in evaluating
the above integral the same infra-red regularization is applied
as in the pure magnetic background.
With the pure electric background
the eigenfunctions of Det K in the long
distance region (i.e., for $k^2 < b$) become anti-causal and
thus unphysical, just like the eigenfunctions under the pure magnetic background
become tachyonic and unphysical in the infra-red
region (i.e., for $k^2 < gH/\sqrt{2})$.
So we must again exclude these
unphysical modes to evaluate the above integral.

The contrast between the effective actions (22) and (42)
is remarkable. First, (42)
has no local minimum. This implies that the electric background
does not generate a condensation. Secondly, (42) has an imaginary part
\bea
Im \thinspace {\cal L}_{g} =-\dfrac{11b^2}{96\pi}.
\eea
This implies that the electric background is unstable. {\it But
perhaps a more important point here is that the imaginary part is negative.
This means that the electric background generates the pair annihilation,
rather than the pair creation, of the gluons. This is because the negative
imaginary part can be interpreted as the negative probability of
the pair creation. This implies that the gluons in QCD, unlike
the electrons in QED, tend to annihilate among
themselves in the color electric field.}
This is really remarkable because this is precisely what one needs
to explain the asymptotic freedom. Remember that the asymptotic freedom
comes from the anti-screening effect, but for this anti-screening one needs
the pair annihilation of gluons in the color electric flux.
This means that our result is not only consistent with the asymptotic
freedom, but actually explains why one must have the asymptotic freedom in QCD.

With this we can now make an independent confirmation
of our effective actions (22) and (42). To do this first notice that
the imaginary part (32) of the Savvidy-Nielsen-Olesen action as well as
ours (22) and (42) are quadratic in the background fields.
In our notation (38) this means that
the imaginary part of the one loop effective action is second
order in the coupling constant $g$. So one can find the
correct imaginary part of the effective action perturbatively,
just by calculating the effective action up to
the second order in the coupling constant in the perturbative
expansion. Now, a remarkable point is that by doing this one
can reproduce our results \cite{schan},
\bea
Im \thinspace \Delta {\cal L}_{g}=\left\{{~~~~0~~~~~~~~~~~~~~ b=0~~
\atop -\dfrac{11b^2}{96\pi}~~~~~~~~~~~a=0~.}\right.
\eea
{\it This confirms that our infra-red regularization is indeed correct.
More importantly this confirms that we do have the desired dynamical
symmetry breaking and the magnetic condensation in QCD.}
In the following we will provide a third
independent argument which supports our results.

An important point to observe here is that the effective actions (22)
and (42) are actually the mirror image of each other.
To see this notice that we can obtain
(42) from (22) simply by replacing $a=gH/\sqrt{2}$ with $-ib$, and similarly
(22) from (42) by replacing $b$ with $ia=igH/\sqrt{2}$.
This is the first indication
that there exists a fundamental symmetry which we call the duality,
in the effective action of QCD. We will discuss this duality in detail
in the following.

The quark loop makes an extra contribution to the effective action.
As we will see in the following we find for the massless quarks,
\bea
\Delta {\cal L}_{q} &=&
-\dfrac{b^2}{48\pi^2} N_f (\ln \dfrac {b}{\mu^2}-c_q)
+ i \dfrac{b^2}{96\pi}, \nn \\
    c_q &=&\gamma +\ln (2 \pi) +\dfrac{6}{\pi^2}\sum_{n=1}^{\infty}
\dfrac{1}{n^2}\ln n.
\eea
So, together with (42), we obtain the following effective action
\bea
&{\cal L}_{eff}=\dfrac{b^2}{2g^2}+\dfrac{11-N_f}{48\pi^2}b^2
(\ln \dfrac{b}{\mu^2}-c_t), \nn\\
&c_t=\dfrac{11}{11-N_f}c_g - \dfrac{1}{11-N_f}c_q,
\eea
which is shown in Fig.2.
\begin{figure}[tbp]
\begin{center}
\epsfig{file= 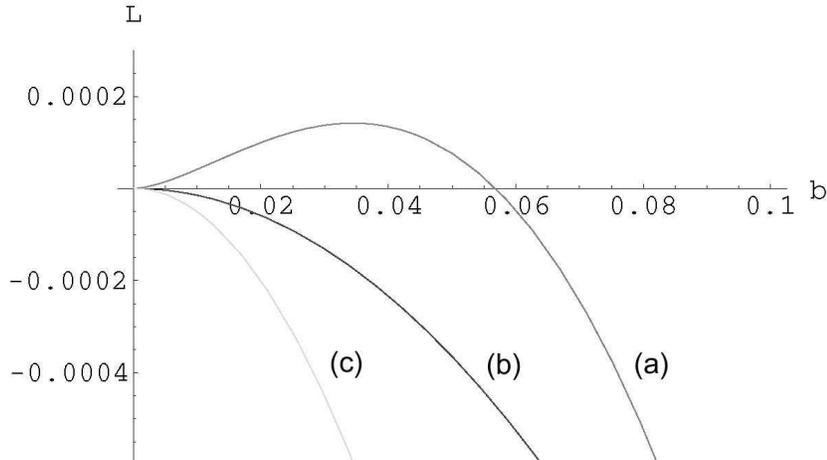, width = 11cm}
\caption[]{The effective action of SU(2) QCD in the pure electrical background.
Here (a) is the real (dispersive) part of the effective action,
(b) is the imaginary (absorptive) part of the effective action,
and (c) is the classical action.}
\end{center}
\end{figure}

\section {Vacuum Stability}

So far we have established the monopole condensation as
a dynamical symmetry breaking. But this by itself does
not allow us to claim that the monopole condensation provides the true physical
vacuum. To show that the monopole condensation is indeed the unique
vacuum of QCD, one must calculate the effective action with an
arbitrary background of the restricted
potential ${\hat A}_{\mu}$ and show that indeed the monopole
condensation provides the true stable minimum of the effective potential.

For a general background with arbitrary $a$ and $b$, the contribution
of the gluon and ghost loops corresponding to the
functional determinant (37) is given by
\bea
&\Delta {\cal L}_{g}  =  \dfrac{1}{16 \pi^2}  \int_{0}^{\infty} \dfrac{ d t}{t^{3-\epsilon}}
 \dfrac{ a b t^2}{\sinh  (a t) \sin (b t) }\nn \\
&\times[\exp(-2at)+\exp(2at)+\exp(2ibt)+\exp(-2ibt)-2]\nn \\
&=\Delta {\cal L}_1 +\Delta {\cal L}_2+\Delta {\cal L}_3,
\eea
where
\bea
&\Delta {\cal L}_1=\dfrac{1}{16 \pi^2}  \int_{0}^{\infty} \dfrac{ d t}{t^{3-\epsilon}}
 \dfrac{ a b t^2 }{\sinh  (a t) \sin (b t) }[\exp(-2at)+\exp(2at)],\nn \\
&\Delta {\cal L}_2=\dfrac{1}{16 \pi^2}  \int_{0}^{\infty} \dfrac{ d t}{t^{3-\epsilon}}
 \dfrac{ a b t^2}{\sinh  (a t) \sin (b t) }[\exp(2ibt)+\exp(-2ibt)],\nn \\
&\Delta {\cal L}_3=-\dfrac{1}{8 \pi^2}  \int_{0}^{\infty} \dfrac{ d t}{t^{3-\epsilon}}
 \dfrac{ a b t^2}{\sinh  (a t) \sin (b t) }.
\eea
Notice that $\Delta {\cal L}_1 +\Delta {\cal L}_2$ describes
the contribution of Det K
of the gluon loop, but $\Delta {\cal L}_3$ describes the contribution of
Det M of the ghost loop. Here again one should keep in mind that the
contour of the integral should pass above the $t$-axis to preserve
the causality.

The integral expression (47) of the effective action has been
known for some time \cite{ditt}, but the actual integration
of it is not easy to perform. Indeed, as far as we understand,
the integration has not been completed satisfactorily.
In the following we will perform the integral, and present
a compact expression of the effective action. To carry out
the integral we need to re-express the integrand in such a way
that we can do the integral analytically. For this purpose we
introduce the following identity \cite{cho5},
\bea
\dfrac{xy}{\sinh (x) \sin(y)} &=& 1- \dfrac{x^2-y^2}{6} -\dfrac{2}{\pi} x^3 y
\sum_{n=1}^{\infty} \dfrac{(-1)^n}{n}\dfrac{{\rm csch}(\dfrac{n\pi y}{x})}{x^2 +n^2
\pi^2}\nn\\
&+&\dfrac{2}{\pi} x y^3 \sum_{n=1}^{\infty}
\dfrac{(-1)^n}{n}\dfrac{{\rm csch}(\dfrac{n\pi x}{y})}{y^2 -n^2 \pi^2}.
\eea
The identity, which we can establish using one of the Ramanujian's
identities\cite{rama}, has played an important role in the calculation of the effective action
of the scalar QED.  Here again in QCD the same identity plays the crucial role in evaluating the effective action.

Now we can calculate $\Delta {\cal L}_{1},~ \Delta {\cal L}_{2}$,
and $\Delta {\cal L}_{3}$ separately with the help of our identity.
Indeed using the identity (49) we obtain the following expression for
$\Delta {\cal L}_{1}$,
\bea
\Delta {\cal L}_1 =I_1(0, 2a)+I_1(0, -2a)+I_2(0, 2a)+I_2(0,-2a)
+ I_3(0, 2a)+ I_3(0, -2a),
\eea
where
\bea
I_1(\epsilon, \lambda)&=&\dfrac{1}{16 \pi^2}  \int_{0}^{\infty}
t^{\epsilon -3} \Big(1- \dfrac{a^2 -b^2}{6} t^2 \Big) e^{-\lambda t}dt\nn\\
&=&\dfrac{1}{16 \pi^2}  \lambda^{-\epsilon}
\Big[\dfrac{\lambda^2}{2}(1+\dfrac{3}{2} \epsilon) -
\dfrac{a^2-b^2}{6}\Big]\Gamma(\epsilon),\nn\\
I_2(\epsilon, \lambda)&=&-\dfrac{ab}{8\pi^2}\sum_{n=1}^{\infty} \dfrac{(-1)^n}{n}{\rm csch}(\dfrac{n\pi b}{a})
 \int_{0}^{\infty} t^{\epsilon +1}\dfrac{e^{-\lambda t}}{t^2 +(\dfrac{n\pi}{a})^2} dt \nn\\
&=&\dfrac{ab}{8\pi^3} \sum_{n=1}^{\infty} \dfrac{(-1)^n}{n}
{\rm csch}(\dfrac{n \pi b}{a})[{\rm ci}(\dfrac{n\pi\lambda}{a})\cos(\dfrac{n\pi\lambda}{a})
+{\rm si}(\dfrac{n\pi\lambda}{a})\sin(\dfrac{n\pi\lambda}{a})],\nn\\
I_3(\epsilon, \lambda)&=&\dfrac{ab}{8\pi^2}\sum_{n=1}^{\infty} \dfrac{(-1)^n}{n}{\rm csch}(\dfrac{n\pi a}{b})
\int_{0}^{\infty} t^{\epsilon +1}\dfrac{e^{-\lambda t}}{t^2 -(\dfrac{n\pi}{b})^2} dt\nn\\
&=&-\dfrac{ab}{16\pi^3} \sum_{n=1}^{\infty} \dfrac{(-1)^n}{n}
{\rm csch}(\dfrac{n \pi a}{b})[{\rm Ei}(-\dfrac{n\pi\lambda}{b})\exp(\dfrac{n\pi\lambda}{a})
+{\rm Ei}(\dfrac{n\pi\lambda}{b})\exp(-\dfrac{n\pi\lambda}{a})].
\eea
Notice that here ci($x$) and si($x$) are the cosine and sine
integral functions, and Ei($-x$) is the exponential integral
function \cite{abra},
\bea
{\rm ci}(x)&=&-\int_{x}^{\infty} \dfrac{\cos(t)}{t} dt \nn\\
&=&\gamma+\ln x +\sum_{n=1}^{\infty} \dfrac{(-1)^n x^{2n}}{(2n)!2n},
~~~~~~~~~~({\rm Re}~x>0)\nn\\
{\rm si}(x)&=&-\int_{x}^{\infty} \dfrac{\sin(t)}{t} dt \nn\\
&=&-\dfrac{\pi}{2}-\sum_{n=1}^{\infty} \dfrac{(-1)^n x^{2n-1}}{(2n-1)!(2n-1)},
~~~~~({\rm Re}~x>0)\nn\\
{\rm Ei}(-x)&=&-\int_{x}^{\infty} \dfrac{e^{-t}}{t} dt \nn\\
&=&\gamma+\ln x +\sum_{n=1}^{\infty} \dfrac{(-1)^n x^{n}}{(n)!n}.
~~~~~~~~~~~~({\rm Re}~x>0)
\eea
With these we find
\bea
\Delta {\cal L}_1&=&\dfrac{11a^2+b^2}{48\pi^2}(\dfrac{1}{\epsilon}-\gamma)+\dfrac{3a^2}{8\pi^2}
-\dfrac{11a^2+b^2}{48\pi^2}\ln(\dfrac{2a}{\mu^2})\nn\\
&+&\dfrac{ab}{4\pi^3} \sum_{n=1}^{\infty} \dfrac{(-1)^n}{n}
{\rm csch}(\dfrac{n\pi b}{a}){\rm ci}(2n\pi)\nn\\
&-&\dfrac{ab}{8\pi^3} \sum_{n=1}^{\infty} \dfrac{(-1)^n}{n}
{\rm csch}(\dfrac{n \pi a}{b})\Big[{\rm Ei}(-\dfrac{2n\pi a}{b})\exp(\frac{2n\pi a}{b})\nn\\
&+&\bar{{\rm Ei}}(\dfrac{2n\pi a}{b})\exp(-\frac{2n\pi a}{b})\Big]+i\dfrac{a^2}{96\pi}\nn\\
&-&i\dfrac{ab}{8\pi^2} \sum_{n=1}^{\infty} \dfrac{(-1)^n}{n}
\Big[{\rm csch}(\dfrac{n\pi b}{a})
+{\rm csch}(\dfrac{n\pi a}{b})\exp(-\frac{2 n\pi a}{b})\Big],
\eea
where $\bar{\rm Ei}(x)$ is $ {\rm Re ~Ei}(x)$.
To obtain the above result notice that, for $\lambda = -2a$, a naive integration
of (51) gives an infra-red divergence. So one must perform
the integral with a positive $\lambda$ first, and then make an analytic
continuation to $\lambda = -2a$. In this analytic continuation one
must keep in mind two points. First, the analytic continuation
should preserve the causality. This means that we must select the correct
(i.e., physical) branch in the analytic continuation of ci($x$), si($x$),
and Ei($-x$) to the negative real axis.
Secondly, the continuation must
be done in such a way that the unphysical modes should have no contribution.
This requires that the imaginary component of
$\Delta {\cal L}_{1}$ should reproduce the previous result (21)
in the pure magnetic background.
With these precautions we obtain the above result.

Now it is simple to evaluate $\Delta {\cal L}_{2}$, because it can
be put into the same form as $\Delta {\cal L}_{1}$ with a Wick rotation
of $t$ to $it$. Indeed we have (after the Wick rotation)
\bea
\Delta {\cal L}_2=-\dfrac{1}{16 \pi^2}  i^{\epsilon}\int_{0}^{\infty} \dfrac{ d t}{t^{3-\epsilon}}
 \dfrac{ a b t^2}{\sinh  (b t) \sin (a t) }[\exp(-2bt)+\exp(2bt)].
\eea
But an important point to notice here is that with the Wick rotation
the contour of the above integral should now pass below the $t$-axis.
With this observation we obtain
\bea
\Delta {\cal L}_2=&-&\dfrac{11b^2+a^2}{48\pi^2}(\dfrac{1}{\epsilon}-\gamma)-\dfrac{3b^2}{8\pi^2}
+\dfrac{11b^2+a^2}{48\pi^2}\ln(\dfrac{2b}{\mu^2})\nn\\
&-&\dfrac{ab}{4\pi^3} \sum_{n=1}^{\infty} \dfrac{(-1)^n}{n}
{\rm csch}(\dfrac{n\pi a}{b}){\rm ci}(2n\pi)\nn\\
&+&\dfrac{ab}{8\pi^3} \sum_{n=1}^{\infty} \dfrac{(-1)^n}{n}
{\rm csch}(\dfrac{n \pi b}{a})\Big[{\rm Ei}(-\dfrac{2n\pi b}{a})\exp(\frac{2n\pi b}{a})\nn\\
&+&\bar{{\rm Ei}}(\dfrac{2n\pi b}{a})\exp(-\frac{2n\pi b}{a})\Big]
-i\dfrac{11b^2}{96\pi}\nn\\
&+&i\dfrac{ab}{8\pi^2} \sum_{n=1}^{\infty} \dfrac{(-1)^n}{n}
\Big[{\rm csch}(\dfrac{n\pi a}{b})-{\rm csch}(\dfrac{n\pi b}{a})\exp(-\frac{2 n\pi b}{a})\Big].
\eea
Again we emphasize that the above Wick rotation prescription
automatically and naturally guarantees that we have the same infra-red
regularization in the evaluation of $\Delta {\cal L}_1$ and $\Delta {\cal
L}_2$. Indeed we can easily confirm that $\Delta {\cal L}_{2}$
reproduces the previous result (41)
in the pure electric background (i.e., in the limit $a$ goes
to zero), as it should.

Finally it is straightforward to calculate $\Delta {\cal L}_{3}$,
because it has no infra-red divergence. We find
\bea
\Delta {\cal L}_3&=&-\dfrac{1}{8 \pi^2}  \int_{0}^{\infty} \dfrac{ d t}{t^{3-\epsilon}}
 \dfrac{ a b t^2}{\sinh  (a t) \sin (b t) }\nn\\
 &=&-2 \Big(I_1(0,0)+ I_2(0,0)+ I_3(0,0)\Big).
\eea
From this we have
\bea
&\Delta {\cal L}_3=\dfrac {a^2-b^2}{48\pi^2 } (\dfrac {1}{\epsilon} - \gamma)
-\dfrac{ab}{4\pi^3} \sum_{n=1}^{\infty} \dfrac{(-1)^n}{n}
\Big[{\rm csch}(\dfrac{n\pi b}{a}) \Big (\ln(\dfrac{n\pi\mu^2}{a})+
\gamma \Big )\nn\\
&-{\rm csch}(\dfrac{n\pi a}{b}) \Big (\ln(\dfrac{n\pi\mu^2}{b})
+ \gamma \Big ) \Big]
+i\dfrac{ab}{8\pi^3} \sum_{n=1}^{\infty} \dfrac{(-1)^n}{n}{\rm csch}(\dfrac{n\pi
a}{b}).
\eea
To obtain the above result we have used the following identity \cite {cho5},
\bea
\dfrac{12}{\pi}ab \sum_{n=1}^{\infty} \dfrac{(-1)^n}{n}
\Big[{\rm csch}(\dfrac{n\pi b}{a})-{\rm csch}(\dfrac{n\pi
a}{b})\Big]=b^2-a^2.
\eea
In evaluating the integrals it should be stressed again that
one must be very careful to implement a proper infra-red regularization.
In particular, one has to make sure that the above result reproduces
the results of the previous sections for the pure magnetic
and the pure electric backgrounds.

With the above results we finally obtain (after the modified
minimal subtraction)
\bea
\Delta {\cal L}_{g}&=&\Delta {\cal L}_1 +\Delta {\cal L}_{2}+\Delta {\cal
L}_{3}\nn\\
&=&\dfrac{3a^2-3b^2}{8\pi^3}-\dfrac{11a^2+b^2}{48\pi^2}\ln
(\dfrac{2a}{\mu^2})+\dfrac{11b^2+a^2}{48\pi^2}\ln (\dfrac{2b}{\mu^2})\nn\\
&+&\dfrac{ab}{4\pi^3} \sum_{n=1}^{\infty} \dfrac{(-1)^n}{n}
\Big({\rm csch}(\dfrac{n\pi b}{a})-{\rm csch}(\dfrac{n\pi a}{b})\Big)({\rm ci}(2n\pi)-\gamma)\nn\\
&+&\dfrac{ab}{8\pi^3} \sum_{n=1}^{\infty} \dfrac{(-1)^n}{n}\Big[{\rm csch}(\dfrac{n\pi b}{a})
\Big(\exp(\frac{2n\pi b}{a}){\rm Ei}(-\dfrac{2n\pi b}{a})\nn\\
&+&\exp(-\frac{2n\pi b}{a})\bar{{\rm Ei}}(\dfrac{2n\pi
b}{a})\Big)-{\rm csch}(\dfrac{n \pi a}{b})\Big(\exp(\frac{2n\pi a}{b}){\rm Ei}(-\dfrac{2n\pi a}{b})\nn\\
&+&\exp(-\frac{2n\pi a}{b})\bar{{\rm Ei}}(\dfrac{2n\pi a}{b})\Big)\Big]\nn\\
&-&\dfrac{ab}{4\pi^3} \sum_{n=1}^{\infty} \dfrac{(-1)^n}{n}
\Big({\rm csch}(\dfrac{n\pi b}{a})\ln(\dfrac{n\pi\mu^2}{a})-{\rm csch}(\dfrac{n\pi a}{b})\ln(\dfrac{n\pi\mu^2}{b})\Big)\nn\\
&-& i\dfrac{a^2+11b^2}{96\pi}-i\dfrac{ab}{8\pi^2} \sum_{n=1}^{\infty} \dfrac{(-1)^n}{n}
\Big({\rm csch}(\dfrac{n\pi b}{a}) \exp(-\frac{2n\pi b}{a})\nn\\
&+&{\rm csch}(\dfrac{n\pi a}{b})\exp(-\frac{2 n\pi a}{b})\Big)+i\dfrac{ab}{8\pi^2} \sum_{n=1}^{\infty} \dfrac{(-1)^n}{n}
{\rm csch}(\dfrac{n\pi a}{b}).
\eea
Notice again that here we have used  our identity (58) to obtain
the above result.
The result is summarized in Fig.3 and Fig.4, where we have
plotted the gluon contribution of the dispersive and absorbtive parts of the effective action.
\par
An important point here is that the effective action acquires
the following imaginary component,
\bea
Im \Delta {\cal L}_{g}=&-&\dfrac{a^2+11b^2}{96\pi} -\dfrac{ab}{8\pi^2} \sum_{n=1}^{\infty} \dfrac{(-1)^n}{n}
\Big({\rm csch}(\dfrac{n\pi b}{a}) \exp(-\frac{2 n\pi b}{a})\nn\\
&+&{\rm csch}(\dfrac{n\pi a}{b})\exp(-\frac{2 n\pi a}{b})\Big)+\dfrac{ab}{8\pi^2} \sum_{n=1}^{\infty} \dfrac{(-1)^n}{n}
{\rm csch}(\dfrac{n\pi a}{b}).
\eea
We can confirm that this expression reproduces the previous results.
Indeed we find
\bea
Im \thinspace \Delta {\cal L}_{1}=\left\{{~~~~0~~~~~~~~~~~~~~ b=0~~
\atop ~~~\dfrac{b^2}{96\pi}~~~~~~~~~~~a=0~,}\right.
\nn\\
Im \Delta {\cal L}_{2}=\left\{{~~~~0~~~~~~~~~~~~~~ b=0~~
\atop -\dfrac{11b^2}{96\pi}~~~~~~~~~~a=0~,}\right.
\nn\\
Im \Delta {\cal L}_{3}=\left\{{~~~~0~~~~~~~~~~~~~~ b=0~~
\atop -\dfrac{b^2}{96\pi}~~~~~~~~~~~a=0~,}\right.
\eea

\begin{figure}[tbp]
\begin{center}
\epsfig{file= 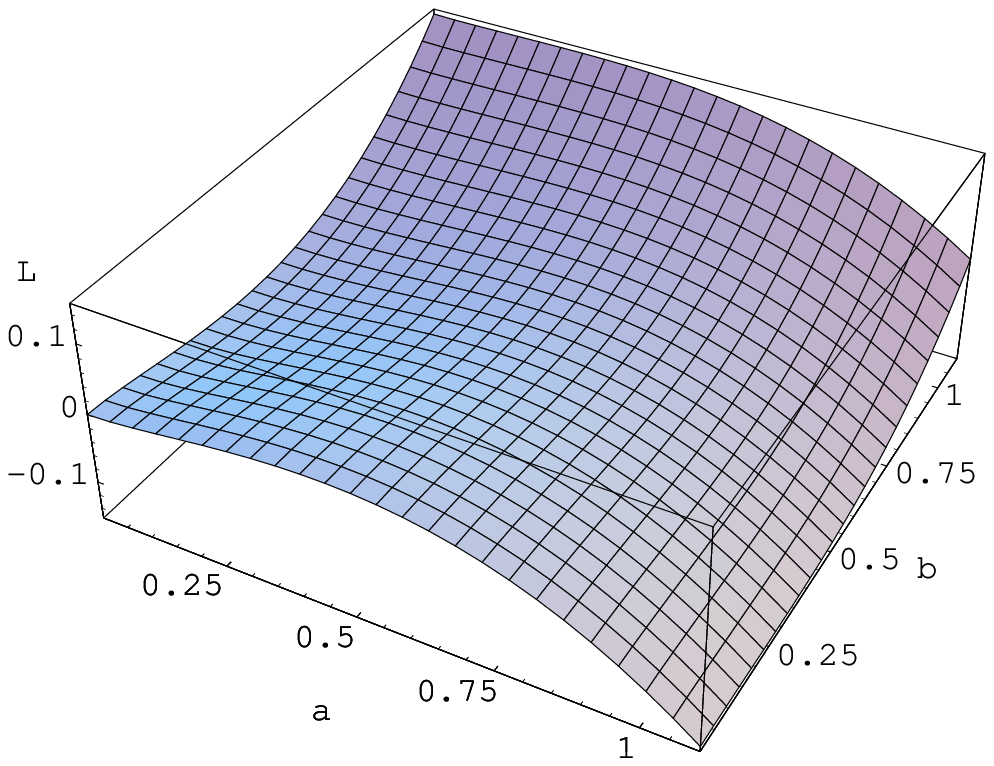, width = 11cm}
\caption[]{The real (dispersive) part of the gluon contribution to the effective action of SU(2) QCD.}
\end{center}
\end{figure}
\begin{figure}[tbp]
\begin{center}
\epsfig{file= 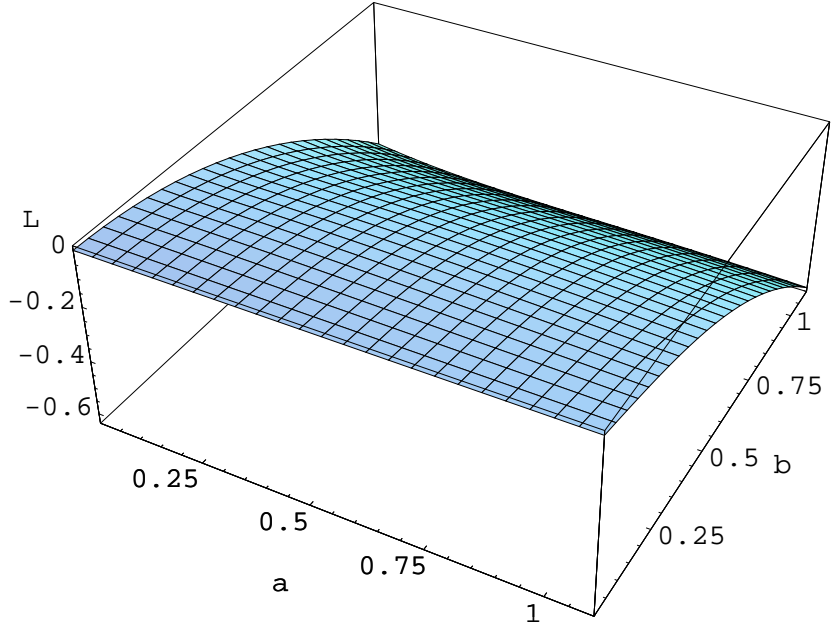, width = 11cm}
\caption[]{The imaginary (absorptive) part of the gluon contribution to the effective action of SU(2) QCD.}
\end{center}
\end{figure}

so that
\bea
Im \Delta {\cal L}_{g}=\left\{{~~~~0~~~~~~~~~~~~~~~ b=0~~
\atop -\dfrac{11b^2}{96\pi}~~~~~~~~~~~a=0~.}\right.
\eea
{\it What is really remarkable about the above result is
the unilateral emergence of the imaginary part with $b \neq 0$.
This immediately tells that the color electric
background generates an instability to the effective action.
Only when the background becomes pure magnetic the imaginary part
disappears completely. This automatically proves that the monopole
condensation is indeed the unique and stable vacuum of QCD, 
at least in one loop approximation.}

Observe that the imaginary part of the effective action
becomes negative in general. This assures again
that the electric background makes the pair annihilation for the gluons,
but not the pair creation. In contrast, as we will see in the following,
the electric background makes the pair creation for the quarks.
This is a very important observation, because this tells that
the gluon pairs behave differently from the quark pairs in the color
electric field. In particular
the valence gluons are not likely to form the glueball bound states.
This explains the experimental fact that there are so few (if at all)
candidates of glueball bound states, while we have towers of
hadronic bound states made of quarks.

\section{Quark Contribution }

Obviously one can not neglect the quarks in QCD.
Let us consider the Lagrangian involving the $SU(2)$
quarks in the fundamental representation
\bea
&{\cal L}_{q} = \bar \Psi( i \gamma^\mu D_\mu -m_q) \Psi, \nn\\
& D_\mu = \pro_\mu + \dfrac{ g}{2i} {\vec \sigma}\cdot \vA_\mu,
\eea
where $m_q$ is the mass of the quarks. One can express the quark
contribution to the effective action
in one loop approximation by
\bea
\Delta {\cal L}_{q}=-\dfrac{ab}{16 \pi^2} N_f \int_{0}^{\infty} \dfrac{dt}{t^{1-\epsilon}}
\coth (at)\cot (bt)\exp(-m_q^2t)  ,
\eea
Notice that formally this is very much like the
well-known expression of the one loop contribution of electron
to the effective action in QED  \cite {schw,cho5}.
This is because at one loop level
only the interaction of the quarks with the restricted potential
contributes to the effective action.

We can evaluate the above integral exactly the same way as we
calculate the electron loop contribution in QED \cite{cho5}.
Using the following Sitaramachandrarao's
identity \cite {cho5,rama},
\bea
xy \coth(x)\cot(y)&=&1+ \dfrac{x^2-y^2}{3} -\dfrac{2}{\pi} x^3 y
\sum_{n=1}^{\infty} \dfrac{1}{n}\dfrac{{\rm csch}(\dfrac{n\pi y}{x})}{x^2 +n^2
\pi^2}\nn\\
&+&\dfrac{2}{\pi} x y^3 \sum_{n=1}^{\infty} \dfrac{1}{n}\dfrac{{\rm csch}(\dfrac{n\pi x}{y})}{y^2 -n^2
\pi^2},
\eea
we obtain (with the modified minimal subtraction),
\bea
\Delta {\cal L}_{q} &=& \dfrac{a^2 - b^2}{48 \pi^2} \ln \dfrac{m_q^2}{\mu^2}\nn \\
&-&\dfrac{ab}{8\pi^3} N_f \sum_{n=1}^{\infty} \dfrac{1}{n}
{\rm coth}(\dfrac{n\pi b}{a})\Big({\rm ci}(\dfrac{2n\pi m_q^2}{a})
\cos(\dfrac{2n\pi m_q^2}{a})
+{\rm si}(\dfrac{2n\pi m_q^2}{a})\sin(\dfrac{2n\pi m_q^2}{a})\Big) \nn\\
&+& \dfrac{ab}{16\pi^3} N_f \sum_{n=1}^{\infty} \dfrac{1}{n}
{\rm coth}(\dfrac{n\pi a}{b})
\Big({\rm Ei} (-\dfrac{2n\pi m_q^2}{b})\exp(\dfrac{2n\pi m_q^2}{b})\nn\\
&+&{\rm Ei}(\dfrac{2n\pi m_q^2}{b})\exp(-\dfrac{2n\pi m_q^2}{b})\Big).
\eea
Notice that the above effective action of the quark loop
also develops an imaginary part
when $b\ne0$,
\bea
Im \thinspace \Delta {\cal L}_q=\dfrac{ab}{16\pi^2} N_f\sum_{n=1}^{\infty}
\dfrac{1}{n}{\rm coth}(\dfrac{n\pi a}{b}) \exp(-\dfrac{2n\pi m_q^2}{b}).
\eea
This is because the exponential integral Ei($-x$) in (66) develops
an imaginary part after the analytic continuation from $x$ to $-x$.
One can compare this with the imaginary part of the gluon loop (60).
Remarkably the signature of the imaginary part of the quark loop is opposite
to that of the gluon loop.  This is due to the opposite statistics between
the gluons and the quarks, which gives
an overall minus sign for the quark loop in (64).
This tells that the quarks should contribute
a positive imaginary part to the effective action when $b\ne0$,
This means that, just like in QED, the electric
flux of the quarks generates the pair
creation and the color screening effect, rather than the pair annihilation
and the color anti-screening effect. This allows the quarks
to form the hadronic bound states.

Observe that in the pure magnetic and pure electric limits 
the above result reduces to
\bea
\Delta {\cal L}_{q}=\left\{{\dfrac{a^2}{48 \pi^2}\ln \dfrac{m_q^2}{\mu^2}
-\dfrac{a^2}{8\pi^4} N_f \sum_{n=1}^{\infty} \dfrac{1}{n^2}
\Big({\rm ci}(\dfrac{2n\pi m_q^2}{a})
\cos(\dfrac{2n\pi m_q^2}{a})\Big)       ~~~~~~~~~~~~~~~~~~b=0,
\atop -\dfrac{b^2}{48 \pi^2}\ln \dfrac{m_q^2}{\mu^2}
+ \dfrac{b^2}{16\pi^4} N_f \sum_{n=1}^{\infty} \dfrac{1}{n^2}
\Big({\rm Ei} (-\dfrac{2n\pi m_q^2}{b})\exp(\dfrac{2n\pi m_q^2}{b})\Big)~~~~~~~~~~a=0.}\right.
\eea
The contribution of the quark loop to the effective action is
plotted in Fig.5 and Fig.6 . In view of the experimental
fact that $\Lambda_{\bar {MS}} \simeq 250$ Mev and $m_q \simeq 5 $ Mev, we have assumed $m_q = 0.02$
to obtain the figures.

Notice that in the massless limit we have\cite{cho5}
\vskip 1cm
\begin{figure}[tbp]
\begin{center}
\epsfig{file= 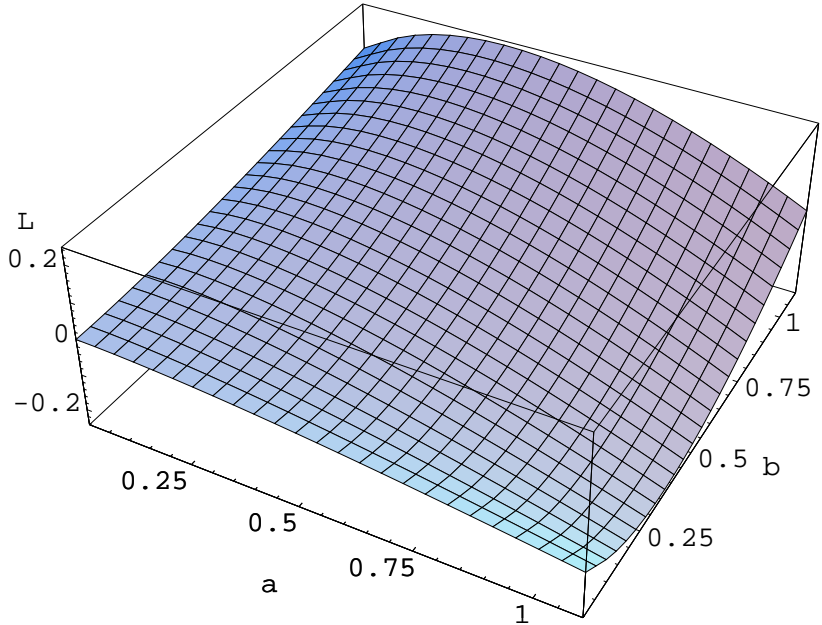, width = 11cm}
\caption[]{The real (dispersive) part of the quark contribution to the effective action of SU(2) QCD.}
\end{center}
\end{figure}
\begin{figure}[tbp]
\begin{center}
\epsfig{file= 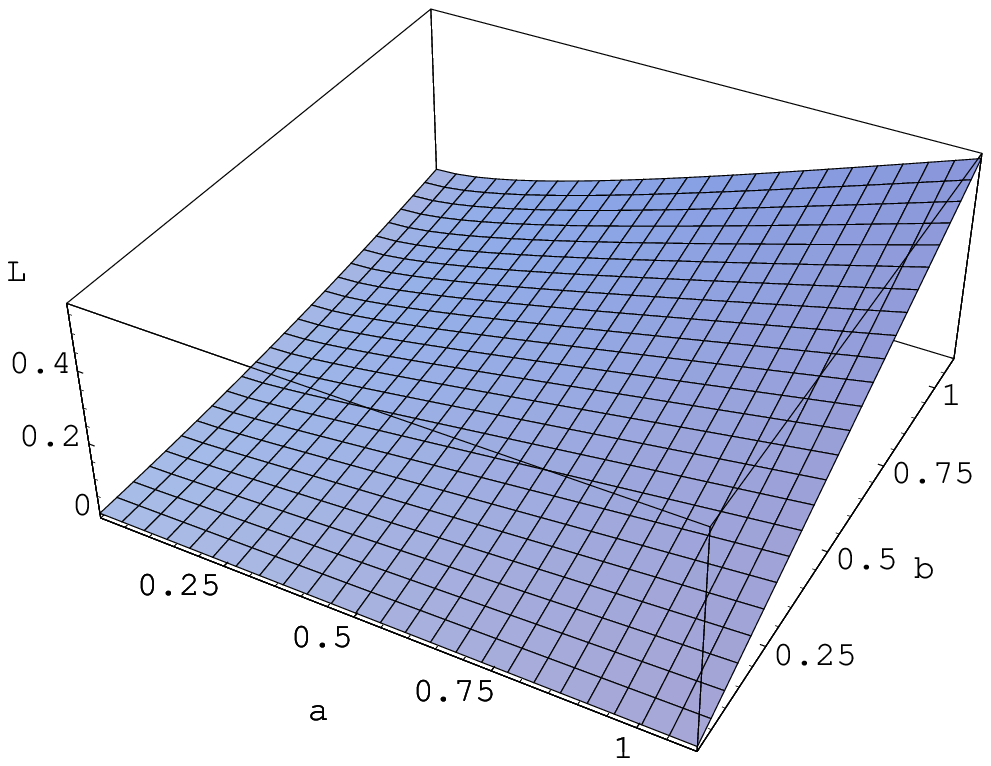, width = 11cm}
\caption[]{The imaginary (absorptive) part of the quark contribution to the effective action of SU(2) QCD.}
\end{center}
\end{figure}

\bea
\Delta {\cal L}_{q} \Big|_{m_q =0} &\simeq&\dfrac{N_f}{48 \pi^2} \Big[ a^2 - b^2 -
\dfrac{6ab}{\pi}\sum_{n=1}^{\infty} \dfrac{1}{n} \Big({\rm coth}(\dfrac{n\pi b}{a})-
{\rm coth} (\dfrac{n\pi a}{b})\Big)\Big] \ln \Big(\dfrac{m_q}{\mu}\Big)^2\nn\\
&-&\dfrac{ab}{8\pi^3}N_f \sum_{n=1}^{\infty} \dfrac{1}{n}
 \Big[{\rm coth} (\dfrac{n\pi b}{a})
\Big(\ln (\dfrac{2n\pi\mu^2}{a})+\gamma\Big)-
{\rm coth} (\dfrac{n\pi a}{b})
\Big(\ln (\dfrac{2n\pi\mu^2}{b})+\gamma\Big)\Big]\nn\\
&+& i\dfrac{ab}{16\pi^2} N_f \sum_{n=1}^{\infty} \dfrac{1}{n}{\rm coth}(\dfrac{n\pi
a}{b}),
\eea
so that, when $ab\ne 0$, the effective action from the quark loop
(unlike the gluon loop) becomes divergent in the massless limit.
This is because here the infra-red divergence comes from the zero
modes which are physical, so that the infra-red divergence can not be
removed.

One could separate the divergent part from the finite part
in $\Delta {\cal L}_{q}$. We find
\bea
\Delta {\cal L}_{q}\Big|_{m_q=0} = \Delta {\cal L}_{\infty}
+ \Delta {\cal L}_{\rm fin} ,
\eea
where
\bea
\Delta {\cal L}_\infty &\simeq& \dfrac{N_f}{48 \pi^2} \Big[ a^2 - b^2 -
\dfrac{6ab}{\pi}\sum_{n=1}^{\infty} \dfrac{1}{n} \Big({\rm coth}(\dfrac{n\pi b}{a})-
{\rm coth} (\dfrac{n\pi a}{b})\Big)\Big] \ln \Big(\dfrac{m_q}{\mu}\Big)^2\nn\\
&+&\dfrac{ab}{16\pi^3} N_f \sum_{n=1}^{\infty}\Big({\rm coth}(\dfrac{n\pi b}{a})+ {\rm coth} (\dfrac{n\pi
a}{b})\Big)\ln \dfrac{a}{b}\nn\\
&+&i\dfrac{ab}{16\pi^2} N_f \sum_{n=1}^{\infty} \dfrac{1}{n}{\rm coth}(\dfrac{n\pi
a}{b}),\nn\\
\Delta {\cal L}_{\rm fin}&=&-\dfrac{ab}{16\pi^2} N_f\sum_{n=1}^{\infty} \dfrac{1}{n}\Big({\rm coth}(\dfrac{n\pi b}{a})-
{\rm coth} (\dfrac{n\pi a}{b})\Big)\nn\\
&& \Big(2 \gamma + \ln
(\dfrac{2n\pi\mu^2}{a})+\ln(\dfrac{2n\pi\mu^2}{b})\Big).
\eea
This tells that one has to keep $m_q$ finite in the evaluation of the
effective action of QCD when $ab\neq0$, to avoid the infra-red divergence of
the quark loop.

But notice that, when $ab=0$, the logarithmic divergence in (69) disappears
due to the following identity \cite{cho5},
\bea
\dfrac{6ab}{\pi} \sum_{n=1}^{\infty} \dfrac{1}{n}\Big({\rm coth}(\dfrac{n\pi b}{a})-
{\rm coth} (\dfrac{n\pi a}{b})\Big)= a^2 -b^2.
\eea
So in this case (i.e., when $ab=0$) $\Delta {\cal L}_{q}$
becomes finite even in the massless limit,
\bea
\Delta {\cal L}_{q}\Big|_{m_q=0}=\left\{{~\dfrac {a^2}{48\pi^2}
N_f(\ln \dfrac {a}{\mu^2} -c_q)~~~~~~~~~~~~~~~~~~~~~~~ b=0
\atop -\dfrac{b^2}{48\pi^2} N_f (\ln \dfrac {b}{\mu^2}-c_q)
+ i \dfrac{b^2}{96\pi} ~~~~~~~~~~~~a=0,}\right.
\eea
where
\bea
c_q = \gamma +\ln(2\pi) + \dfrac{6}{\pi^2}\sum_{n=1}^{\infty}
\dfrac{1}{n^2}\ln n = 2.98504...
\eea
Notice that we can also obtain the above result from (68)
by making the massless limit of the quarks.

\section {Effective Action and Duality}

With the above analysis we can sum up the gluon and quark
contributions and obtain the following
final effective action of $SU(2)$ QCD,
\bea
 {\cal L}_{eff} &=& {\cal L}_0 +\Delta {\cal L}_{g} +\Delta {\cal L}_q  \nn \\
  &=&-\dfrac { a^2 - b^2}{2g^2}+\dfrac{3a^2-3b^2}{8\pi^3}-\dfrac{11a^2+b^2}{48\pi^2}\ln
(\dfrac{2a}{\mu^2})+\dfrac{11b^2+a^2}{48\pi^2}\ln (\dfrac{2b}{\mu^2})
       +\dfrac{a^2 - b^2}{48 \pi^2} \ln \dfrac{m_q^2}{\mu^2}                 \nn\\
&+&\dfrac{ab}{4\pi^3} \sum_{n=1}^{\infty} \dfrac{(-1)^n}{n}
\Big({\rm csch}(\dfrac{n\pi b}{a})-{\rm csch}(\dfrac{n\pi a}{b})\Big)({\rm ci}(2n\pi)-\gamma)\nn\\
&+&\dfrac{ab}{8\pi^3} \sum_{n=1}^{\infty} \dfrac{(-1)^n}{n}\Big[{\rm csch}(\dfrac{n\pi b}{a})
\Big(\exp(\frac{2n\pi b}{a}){\rm Ei}(-\dfrac{2n\pi b}{a})
+\exp(-\frac{2n\pi b}{a})\bar{{\rm Ei}}(\dfrac{2n\pi
b}{a})\Big)\nn\\
&-&{\rm csch}(\dfrac{n \pi a}{b})\Big(\exp(\frac{2n\pi a}{b})
{\rm Ei}(-\dfrac{2n\pi a}{b})
+\exp(-\frac{2n\pi a}{b})\bar{{\rm Ei}}(\dfrac{2n\pi a}{b})\Big)\Big]\nn\\
&-&\dfrac{ab}{4\pi^3} \sum_{n=1}^{\infty} \dfrac{(-1)^n}{n}
\Big({\rm csch}(\dfrac{n\pi b}{a})\ln(\dfrac{n\pi\mu^2}{a})-{\rm csch}(\dfrac{n\pi a}{b})\ln(\dfrac{n\pi\mu^2}{b})\Big)\nn\\
&+&\dfrac{ab}{8\pi^3} N_f \sum_{n=1}^{\infty} \dfrac{1}{n}
{\rm coth}(\dfrac{n\pi b}{a})\Big({\rm ci}(\dfrac{2n\pi m_q^2}{a})
\cos(\dfrac{2n\pi m_q^2}{a})
+{\rm si}(\dfrac{2n\pi m_q^2}{a})\sin(\dfrac{2n\pi m_q^2}{a})\Big) \nn\\
&+& \dfrac{ab}{16\pi^3} N_f \sum_{n=1}^{\infty} \dfrac{1}{n}
{\rm coth}(\dfrac{n\pi a}{b})
\Big({\rm Ei} (-\dfrac{2n\pi m_q^2}{b})\exp(\dfrac{2n\pi m_q^2}{b})
+{\rm Ei}(\dfrac{2n\pi m_q^2}{b})\exp(-\dfrac{2n\pi m_q^2}{b})\Big)\nn\\
&-& i\dfrac{a^2+11b^2}{96\pi}-i\dfrac{ab}{8\pi^2} \sum_{n=1}^{\infty} \dfrac{(-1)^n}{n}
\Big({\rm csch}(\dfrac{n\pi b}{a}) \exp(-\frac{2n\pi b}{a})\nn\\
&+&{\rm csch}(\dfrac{n\pi a}{b})\exp(-\frac{2 n\pi a}{b})\Big)+i\dfrac{ab}{8\pi^2} \sum_{n=1}^{\infty} \dfrac{(-1)^n}{n}
{\rm csch}(\dfrac{n\pi a}{b}) \nn \\
&+&i\dfrac{ab}{16\pi^2} N_f\sum_{n=1}^{\infty}
\dfrac{1}{n}{\rm coth}(\dfrac{n\pi a}{b}) \exp(-\dfrac{2n\pi m_q^2}{b})  .
\eea
From this we finally obtain Fig.7 and Fig.8,  which
describe the real and imaginary parts of the total effective
action of $SU(2)$ QCD.
Remember that we have assumed $\alpha_s =1, \, \mu =1, \, m_q = 0.02,$
and $N_f =2$ to obtain the figures.
\vskip 1cm
\begin{figure}[tbp]
\begin{center}
\epsfig{file= 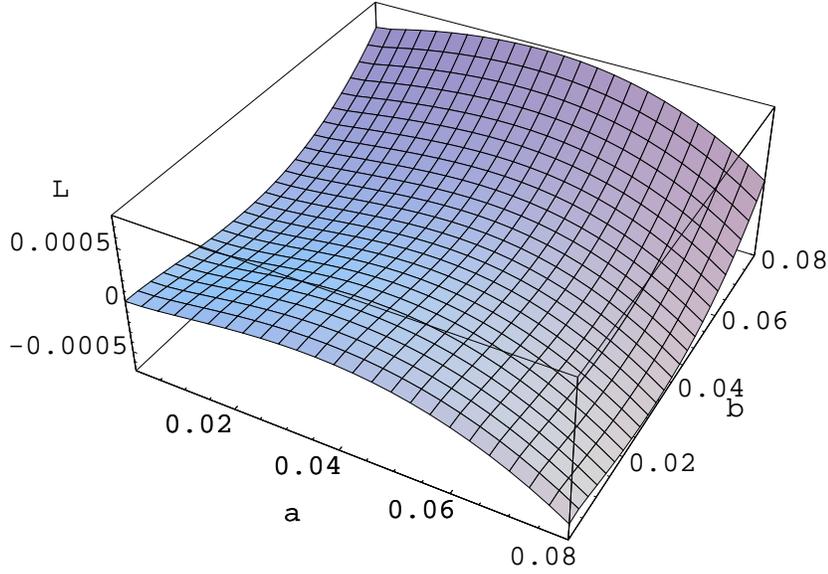, width = 11cm}
\caption[]{The real (dispersive) part of the effective action of SU(2) QCD.}
\end{center}
\end{figure}
\begin{figure}[tbp]
\begin{center}
\epsfig{file= 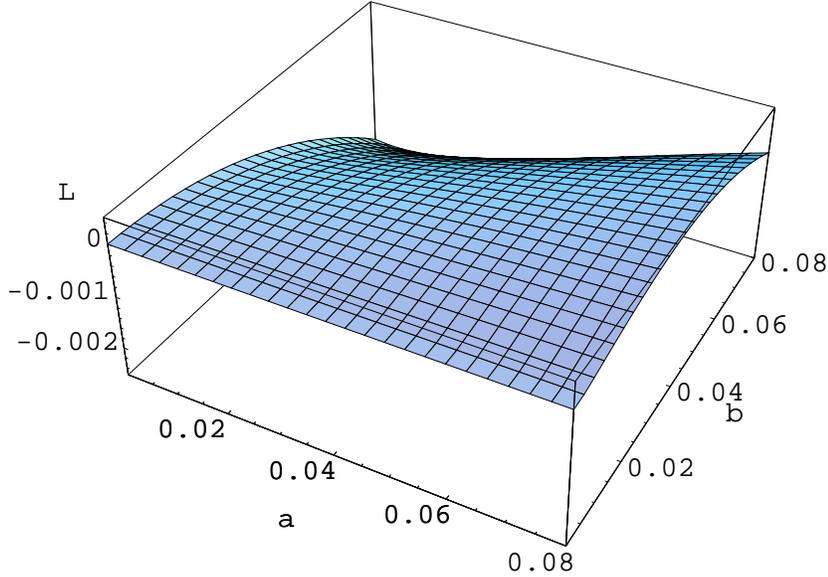, width = 11cm}
\caption[]{The imaginary (absorptive) part of the effective action of SU(2) QCD.}
\end{center}
\end{figure}
\par
{\it A truly remarkable feature of our effective action
is that it
is manifestly invariant under the dual transformation
\bea
a \rightarrow -ib,  ~~~~ b \rightarrow ia.
\eea
In fact
$\Delta {\cal L}_{g}$ and $\Delta {\cal L}_q$
independently can be shown to be invariant
under the dual transformation}.  This tells that,
as a function of $z=a+ib,$
the effective action of QCD is invariant under
the reflection from $z$ to $-z$.
To establish the duality in the effective action it is important
to realize that the argument of the special functions ci($x$), si($x$),
and Ei($-x$) changes the signature under the dual transformation. So the dual transformation
automatically involves the analytic continuation of $x$ to $-x$,
and one must figure out how to make the correct analytic continuation
under the dual transformation. Here again the causality
becomes the guiding principle. Observe that the causality
requires that under the dual transformation we have
\bea
a-i\epsilon&&~\longrightarrow~-i(b+\epsilon)\nn\\
b+\epsilon&&~\longrightarrow~ i(a-i\epsilon),
\eea

so that we must have
\bea
\dfrac{a}{b}-i\epsilon&&~\longrightarrow~-\dfrac{b}{a}-i\epsilon\nn\\
\dfrac{b}{a}+i\epsilon&&~\longrightarrow ~-\dfrac{a}{b}+i\epsilon.
\eea
With this it is straightforward to establish that each of
$\Delta {\cal L}_1 +\Delta {\cal L}_2,~ \Delta {\cal L}_3$,
and $\Delta {\cal L}_q$ separately is invariant under the duality.

From the physical point of view the
existence of the duality is not so surprising. In fact one
should have expected this, because the integral expression (47) of the
effective action evidently has this duality. The really remarkable fact is
that this duality is borne out from our calculation of the effective action.
It must be emphasized that this is a non-trivial feat, because
the effective actions from the earlier calculations
(including the Savvidy-Nielsen-Olesen effective action) have no such duality.
This means that the duality provides a powerful tool to check the consistency of
the one loop effective action. In particular one can use the duality
to check the imaginary part of the effective action, because
the duality intrinsically involves the analytic continuation, and
thus mixes the real and imaginary parts of the effective action
in a non-trivial manner. {\it We emphasize that the consistency with
the duality of our effective action provides
another (a third) independent argument which supports our results.}

Notice that this type of electric-magnetic duality has also been
established recently in the effective action of QED \cite{cho5}. This tells
that our duality is a generic feature of the gauge theories,
both Abelian and non-Abelian.

It should be emphasized that it is exactly the same interaction which provided
the asymptotic freedom that is responsible for the confinement.
The underlying dynamics for both the asymptotic freedom and
the magnetic confinement is the magnetic moment interaction. In this
interaction the only difference between the quark and
the gluon is the color charge, the
gyromagnetic ratio, and the statistics.
It has been well known that the gluon contributes positively but the quark
contributes negatively to the asymptotic freedom. In this paper
we have argued that this was because the gluon
generates the anti-screening effect but the quark generates the
screening effect. Furthermore we have proved that exactly the same
physics ensures the monopole condensation and the confinement.
A simple consequence of this
is that we need exactly
the same maximum number of the quark flavor, $N_f=10$ for $SU(2)$ and
$N_f = 16$ for $SU(3)$, to guarantee both
the asymptotic freedom and the confinement.

Notice that with (23)
the effective Lagrangian can be approximated
as
\bea
{\cal L}_{eff}\simeq&-&\frac14\vec{H}_{\mu\nu}^2-\frac12m^2\vec{C}_\mu^2\nn\\
=&-&\frac{m^2}{2g^2}(\partial_\mu \hat{n})^2-\frac{1}{4g^2}(\partial_\mu \hat{n}
\times\partial_\nu \hat{n})^2,
\eea
near the trivial vacuum.
This of course is nothing but the Skyrme-Faddeev Lagrangian which allows the
topological knot solitons as the classical solutions \cite {faddeev}.
This shows that there exists a deep connection between the generalized non-linear
sigma model and QCD, which is very interesting.

\section {Discussion}

In this paper we have demonstrated the existence of a genuine
dynamical symmetry breaking in QCD triggered by
the monopole condensation. Furthermore we have established that
the monopole condensation describes the stable unique vacuum of QCD. We were
able to do this by calculating the one loop effective action of $SU(2)$
QCD. There have been earlier attempts to calculate the effective
action, but these attempts have not produced a satisfactory result.
We have obtained a compact expression of the effective
action with an arbitrary background field. In the special cases
in which the compact expressions of the effective action
were available (in particular in the pure magnetic background),
our result differs from the earlier results. The main
difference with the earlier attempts was the controversial imaginary
part in the effective action in the pure magnetic background.
This has made the Savvidy-Nielsen-Olesen vacuum unstable. This
assertion on the instability of the vacuum has never been
seriously challenged, nor convincingly revoked.
Our analysis tells that this assertion
is based on the improper infra-red
regularization in the evaluation of the effective action, as
Schanbacher first argued \cite {schan}. Indeed with a proper
infra-red regularization we have shown that the QCD vacuum is
not only stable, but is unique, made of the monopole condensation.
We have provided three
independent arguements to support our conclusion.

It is truly  remarkable (and surprising) that the principles of the  quantum
field theory allow us to demonstrate the confinement
within the framework of QCD.
This appears against the
conventional wisdom. Recently increasing number
of people have been questioning the ability
of the quantum field theory to
provide the confinement in QCD. Indeed
the failure to establish the confinement
within the framework of QCD has encouraged the idea that perhaps
a supersymmetric generalization of QCD may be necessary to ensure
the confinement \cite {witt}. Our analysis shows that
this is not necessary after all. The QCD by itself is able to generate the confinement.
{\it What made this possible for us is the
realization that we must treat the tachyonic bound states
in the magnetic background and the anti-causal propagating states
 in the electric background which exist in the long
distance region as the unphysical modes, and exclude them
from the physical spectrum. In particular, we must exclude
these unphysical modes from the calculation of the effective action with a proper infra-red
regularization}. Only this exclusion of the
unphysical modes can give us a consistent theory of QCD.
The fact that one could establish the dynamical symmetry breaking
and the confinement in QCD within the framework of the existing
quantum field theory should be interpreted as a triumph,
indeed a most spectacular triumph, of the quantum field theory itself.

We conclude with the following remarks: \\
1) It should be emphasized that our analysis is based on the
gauge independent decomposition (1) of the non-Abelian gauge potential to
the restricted  potential $\hat {A}_{\mu}$ and the
valence gluon
$\vec {X}_{\mu}$. This is made possible
with our Abelian projection \cite {cho1,cho2}. The restricted  potential
satisfies the full
non-Abelian gauge degrees of freedom  and forms a non-Abelian
connection space of its own, in spite of the fact that it describes
only the dual dynamics. The valence gluon forms
a gauge covariant vector field, and has the gauge
invariant magnetic moment interaction (10) with the restricted  potential.
And it is this interaction that is responsible for both
the asymptotic freedom and the confinement.
The existence of the gauge independent decomposition
of the non-Abelian potential and a self-consistent restricted QCD
has been known for more than twenty years \cite {cho1,cho2},
but its physical significance
appears to have been appreciated very little so far.
Now we emphasize that it is this decomposition
which allows us to obtain the effective action of QCD.
In particular, it is this decomposition which shows that the vacuum
condensation is indeed made of the monopole condensation.
Many of the earlier approaches had the critical defect that the
decomposition of the non-Abelian gauge potential to the $U(1)$ potential
and the charged vector field
was not gauge independent, which has made these approaches controversial. \\
2) One might question (legitimately)
the validity of the one loop approximation,
since in the infra-red limit the non-perturbative effect
is supposed to play the essential role
in QCD. Our attitude on this issue is that QCD can be viewed as the
perturbative extension of the topological field theory described
by the restricted QCD, so that the non-perturbative
effect in the low energy limit can effectively be represented by
the topological structure of the restricted gauge
theory. This is reasonable,
because the large scale structure of the monopole topology
naturally describes the long range behavior
of the theory. In fact one can argue that it is the
restricted potential that contributes to the Wilson loop integral,
which provides a natural confinement criterion in QCD
\cite{cho3}.
So we believe that our  monopole
background automatically
takes care of the essential feature of the non-perturbative effect.
Of course, one could go further and try to calculate the two loop
effective action \cite{sato}, which certainly will improve  our one loop
correction. But this improvement is
not expected to give any qualitative change, so that
the generic  features of the one loop effective action
and the underlying physics will remain the same. \\
3) There have been two competing proposals for the correct mechanism
of the confinement in QCD, the one emphasizing the role of the instantons and
the other emphasizing that of the monopoles. Our analysis strongly
favors the monopoles as the physical
source for the confinement. It provides a natural dynamical symmetry
breaking, and generates the mass
gap necessary for the confinement in QCD.
Notice that the multiple  vacua, even though it is an important
characteristics of the restricted gauge theory, did not play any crucial role
in our calculation of the effective action. Moreover our result shows that it is
the monopole condensate, not the $\theta$-vacuum, which describes
the physical vacuum of QCD.

Although we have concentrated to $SU(2)$ QCD in this paper, it must be clear
from our analysis that the magnetic condensation is a generic
feature of the non-Abelian gauge theory.
A more detailed discussion which
supports our conclusions and the generalization of
our result to $SU(3)$
will be presented in a forthcoming paper \cite {cho6}.\par
\vspace{0.5cm}\hspace{-0.5cm}
{\bf Acknowledgements}\par
\vspace{0.5cm}
One of the authors (YMC) thanks S. Adler, L. Faddeev, and A. Niemi
for the fruitful discussions, and Professor C. N. Yang for
the continuous encouragements. The other (DGP)
thanks Professor C. N. Yang for the fellowship at Asia Pacific
Center for Theoretical Physics, and appreciates Haewon Lee
for numerous discussions.
The work is supported in part
by the BK21 project of Ministry of Education.

%%%%%%%%%%%%%%%%%%%%%%%%%%%%%%%%%%%%%%%
%%                                                   %%%%%%%%%%%%%%
%%    the bibliography environment   %%%%%%%%%%%%%%
%%                                                    %%%%%%%%%%%%%%
%%%%%%%%%%%%%%%%%%%%%%%%%%%%%%%%%%%%%%%
%%%%%%%%%%%%%%%%%%%%%%%%%
\end{document}